\documentclass[a4paper,12pt]{article}

\usepackage[left=2cm,top=2cm,right=2cm,bottom=2cm,nohead]{geometry}
\usepackage[format=plain,margin=0.5cm,textformat=simple,justification=justified,small,parskip=0cm]{caption}
\usepackage[pdftex]{graphicx}
\usepackage[pdftex,unicode,pdfstartview=FitH,colorlinks=true,pdfborder={0 0 0 0},urlcolor=blue]{hyperref}
\usepackage{amsmath}
\usepackage{cite}

\newcommand{\rmd}{\mathrm{d}}

\title{Control of the socio-economic systems using herding interactions}
\author{A. Kononovicius, V. Gontis}

\date{}

\begin{document}

\maketitle

\begin{center}
\footnotesize Vilnius University, Institute of Theoretical Physics and Astronomy
\end{center}

\begin{abstract}
Collective behavior of the complex socio-economic systems is heavily influenced by
the herding, group, behavior of individuals. The importance of the herding behavior may
enable the control of the collective behavior of the individuals. In this contribution we consider
a simple agent-based herding model modified to include agents with controlled state. We show that
in certain case even the smallest fixed number of the controlled agents might be enough to control
the behavior of a very large system.

\end{abstract}

\section{Introduction}

Collective behavior observed in many complex systems cannot be understood as a simple sum or average over
the behavior of the individual interacting parts \cite{Waldrop1992SAS}. When considering complex socio-economic
systems it is irresistible to see the endogenous interactions behind
the spontaneous emergence of trends, norms or even mass panic. Such phenomena, especially panic, cannot
simply emerge from the rational representative agent framework as the agent is assumed to act completely rationally
\cite{Akerlof2009Princeton,Bouchaud2013JStatPhys,Kirman2010Routledge}. Thus the contemporary
socio-economic research needs to use a different framework to understand these phenomena better \cite{Axelrod1997Comp,Bouchaud2008Nature,Conte2012EPJ,Cristelli2012Fermi,Farmer2012EPJ,Havlin2012EPJ}.

One of the suitable alternative frameworks is heterogeneous agent-based modeling \cite{Conte2012EPJ,
Cristelli2012Fermi,Farmer2012EPJ}. This framework uses a generalized concept of the agent to represent the
interacting parts of the modeled complex system. Interactions between them usually follow very
simple rules, by the virtue of the agents' zero-intelligence or bounded rationality assumption.
Such assumptions can be viewed just as a result of statistical irrelevance of a more detailed consideration. 
Despite the underlying simplicity, the complex collective behavior emerges as a result of the interactions
\cite{Bouchaud2013JStatPhys,Kirman2010Routledge,Cristelli2012Fermi,Alfi2009EPJB1,Alfi2009EPJB2,Chakraborti2011RQUF2,
Feng2012PNAS,Frederick2013PNAS,Kaizoji2005Springer,Kononovicius2012PhysA,Lye2012PhysA,Lux1999Nature,Ruseckas2011EPL}.
One of the main ingredients of these simple rules and emergence of complex collective behavior is
imitation, peer pressure and strong coupling \cite{Bouchaud2013JStatPhys,Kirman2010Routledge,Cristelli2012Fermi,Ruseckas2011EPL}.

Imitation, peer pressure and strong coupling between the agents may allow a possibility for a small fraction of the agents
to make a significant impact on the collective behavior. 
The influence of the small number of individuals on the collective behavior of a crowd
was studied in a series of experiments by Dyer et al.
\cite{Dyer2009RSTB}. People participating in these experiments were asked to move randomly,
but to stay with a crowd. Some of the people in a crowd, a small number of them, were asked
to move in a certain direction. It was expected that they will be able to lead the whole crowd in that direction.
The results of the experiment have shown that $4-10$ directed individuals were enough
to lead the crowds of up to $200$ people. It is interesting to note that the necessary number of directed
individuals grows slower than the total number of people in the crowd. Consequently
the movement of even larger crowds could be also controlled in a similar fashion without a further
significant increase in a total number, not percentage, of the directed individuals. In the context
of this contribution we could see the directed individuals in the aforementioned experiment as
the controlled individuals. Similar experiments were preformed with animals by using controlled
robots \cite{Krause2011TEcoEvo}.

From a point of view of mathematical modeling a similar idea was previously tested in the well-known
Prisonner's dilemma setup by Schweitzer et al. \cite{Schweitzer2012CoRR}. The model was setup
in a way to show that the herding behavior may enhance cooperative behavior instead of a
more self-interested behavior.

We approach the modeling of the collective behavior control slightly differently. In this contribution
we consider Kirman's agent-based model \cite{Kirman1993QJE}. This simple model describes the two
state system dynamics, where agents make decisions based on the individual preferences and
herding. In the recent years an interplay of a different types of social behavior were
broadly studied by the researchers from very different fields \cite{Castellano2009RevModPhys},
yet we feel that Kirman's model is one of the simplest mathematical models for the social behavior.
Consequently we will use Kirman's model to demonstrate the influence of the individuals with a fixed
opinion, which does not change due to the endogenous activity, but does change only due to the
exogenous factors. We will show that this influence may be used to control the behavior of
a social system.

In the Section \ref{sec:kirman} we will present a more detailed discussion on the Kirman's agent-based
herding model and its macroscopic treatment, which was previously done in \cite{Kononovicius2012PhysA,Alfarano2005CompEco,Alfarano2008Dyncon}. In the following section, Section \ref{sec:control}, we will deal with the introduction
of the controlled agents and discuss their effect on the collective behavior of the system. And finally
in the last section of this contribution we will provide a brief summary and discussion.

\section{Kirman's agent-based herding model}
\label{sec:kirman}

In \cite{Kirman1993QJE} Kirman pointed out that a group of entomologists and numerous
economists have observed a very similar phenomena in rather different systems.
The group led by Pasteels observed an ant colony with two identical food sources available
\cite{Pastels1987Birkhauser1,Pastels1987Birkhauser2}. At any given time the majority of ants
used only one of the available food sources, though naturally one would expect
that the both food sources would be exploited equally. It was also observed that from
time to time the preferred food source was switched. Interestingly enough
these switches were triggered not by the exogenous forces, but by the system itself. Similarly Becker
\cite{Becker1991JPolitEco} noted that some of the decisions in economical scenarios might also
have a similar nature - humans also tend to act asymmetrically in apparently symmetrical setups.

Having taken the aforementioned, and other (see references of \cite{Kirman1993QJE}), observations
into account Kirman proposed a simple one-step transition model. Which in general case can be expressed
via the following one-step transition probabilities \cite{Aoki2007Cambridge}:
\begin{equation}
p (X \rightarrow X+1) = (N-X) \mu_1(N,X) \Delta t , \quad p (X \rightarrow X-1) = X \mu_2(N,X) \Delta t , \label{eq:onestepprob}
\end{equation}
here $N$ is a fixed number of agents in the system (one of the available states is occupied by $X$ agents and
the other by  $N-X$ agents), while $\mu_i(N,X)$ are the transition rates. The overall transition rates in the Kirman's
model are composed of the idiosyncratic transition rate, $\sigma_i$, and herding behavior, $h$, terms. One can
define the overall transition rates to be given by \cite{Kononovicius2012PhysA,Alfarano2005CompEco,Alfarano2008Dyncon}
\begin{equation}
\mu_1(N,X) = \sigma_1 + h X , \quad \mu_2(N,X) = \sigma_2 + h (N-X) ,\label{eq:nonext}
\end{equation}
or by
\begin{equation}
\mu_1(N,X) = \sigma_1 + \frac{h}{N} X , \quad \mu_2(N,X) = \sigma_2 + \frac{h}{N} (N-X) .\label{eq:ext}
\end{equation}
Which form of the transition rates is more appropriate depends on the interpretation of the Kirman's model. In the first, Eq. (\ref{eq:nonext}),
case it is assumed that all agents may interact with all other agents, or namely on a global scale. While in the second
case, Eq. (\ref{eq:ext}), the agents are assumed to interact only with the fixed number of other agents, or their local
neighborhood. The main
difference between the two forms is a different scaling of the herding induced transition rates. In the first case they
grow linearly together with the system size, $N$, while in the second case they remain constant. The differences
between these forms can be well understood from the point of view of network theory \cite{Alfarano2009Dyncon}.

Note that the one-step transition probabilities, Eq. (\ref{eq:onestepprob}), scale similarly - as $N^2$ and $N$
correspondingly. Thus we will further refer to these interpretations as the non-extensive and extensive. Identical
reasoning and terminology is also used
in the previous works by Alfarano et al. \cite{Alfarano2005CompEco,Alfarano2008Dyncon,Alfarano2009Dyncon}.

The different scaling of the one-step transition probabilities implies the essential difference in the macroscopic
behavior of these two interpretations of Kirman's model. In the non-extensive case the macroscopic dynamics,
for $x=X/N$, (in the limit $N \rightarrow \infty$) are well described by the stochastic
differential equation \cite{Kononovicius2012PhysA,Alfarano2005CompEco}:
\begin{equation}
\rmd x = [ \sigma_1 (1-x) - \sigma_2 x ] \rmd t + \sqrt{2 h x (1-x)} \rmd W , \label{eq:nonextsde}
\end{equation}
where $W$ stands for a standard one dimensional Brownian motion (or alternatively for a Wiener process).
While in the extensive case the fixed transition rates imply that the diffusion term disappears in the limit of large system
sizes, $N \rightarrow \infty$. In such case the macroscopic dynamics are well described by the ordinary differential equation:
\begin{equation}
\rmd x = [ \sigma_1 (1-x) - \sigma_2 x ] \rmd t + \sqrt{\frac{2 h x (1-x)}{N}} \rmd W \approx [ \sigma_1 (1-x) - \sigma_2 x ] \rmd t . \label{eq:extsde}
\end{equation}
Similar macroscopic model was obtained from a point of view of game theory and studied in the large but finite
system size limit \cite{Traulsen2012PhysRevE}. It should be evident that in the case of large but finite system size
one would have a Gaussian-like fluctuations around the deterministic solution of Eq. (\ref{eq:extsde}).

\section{The control of the collective behavior}
\label{sec:control}

Let us now additionally introduce $M$ agents, whose choice of the state is controlled externally, into the 
herding model. Namely unlike the other
agents, the controlled agents do not switch their state due to endogenous interactions, though they are able
to trigger endogenous switches of the other agents.

As we have discussed in the previous section the agents may interact either locally or globally. If the interaction
is local, then the herding terms disappear or become negligible in the limit of large system sizes. In order for
the controlled agents to make a significant impact on a whole macroscopic system
they have to interact on a global scale. In such case we have two sets of the one-step transition probabilities (analogous to Eqs. (\ref{eq:nonext}) and (\ref{eq:ext})):
\begin{eqnarray}
\mu_1(N,X) = \sigma_1 + h (M_1+X) , & \mu_2(N,X) = \sigma_2 + h (N-X+M-M_1) , \label{eq:cnonext} \\
\mu_1(N,X) = \sigma_1 + \frac{h}{N} X + h M_1 , & \mu_2(N,X) = \sigma_2 + \frac{h}{N} (N-X) + h (M-M_1) . \label{eq:cext}
\end{eqnarray}
In the above $M_1$ is a number of the controlled agents ($M_1 \leq M$) in the state which is occupied by $X$ other agents.

From a purely mathematical point of view the influence of the controlled agents can be included into the individual
behavior parameters, $\sigma_i$. Namely, one can set $\tilde{\sigma}_1 = \sigma_1 + h M_1$
and $\tilde{\sigma}_2 = \sigma_2 + h (M-M_1)$  to return to the original form of the herding model with
shifted individual preferences, $\tilde{\sigma}_i$. Similar approaches may be found in
\cite{Alfarano2013EJF,Carro2013}. In \cite{Carro2013} the external forces
are assumed to drive the periodic fluctuations of the herding behavior parameter, while in our case the controlled
agents act upon individual behavior parameters. In \cite{Alfarano2013EJF} a case where small number of core
agents influence behavior of large number of periphery agents is considered, while our approach lacks
strict hierarchy.

The macroscopic dynamics influenced by the controlled agents, in the limit $N \rightarrow \infty$, are given by, for the non-extensive case,
\begin{equation}
\rmd x = [ (\sigma_1+ h M_1) (1-x) - (\sigma_2+ h \{M-M_1\}) x ] \rmd t + \sqrt{2 h x (1-x)} \rmd W , \label{eq:cnonextsde}
\end{equation}
and, for the extensive case,
\begin{equation}
\rmd x = [ (\sigma_1+ h M_1) (1-x) - (\sigma_2+ h \{M-M_1\}) x ] \rmd t . \label{eq:cextsde}
\end{equation}
It should be straightforward to determine the stationary mean of Eq. (\ref{eq:cnonextsde}), $\bar x$,
(the recipe is given in most stochastic calculus handbooks (e.g., \cite{Gardiner2009Springer}))
and the fixed point of Eq. (\ref{eq:cextsde}), $x_f$,
\begin{equation}
\bar x = x_{f} = \frac{h M_1+\sigma_1}{h M+\sigma_1+\sigma_2} . \label{eq:exmean}
\end{equation}
Note that the long term impact of the controlled agents depends only on their number and the strength of
individual preferences of the other agents. So, in this case, one can use a fixed number of the
controlled agents to influence the behavior of an infinitely large system.

\begin{figure}
    \centering
    \includegraphics[width=0.4\textwidth]{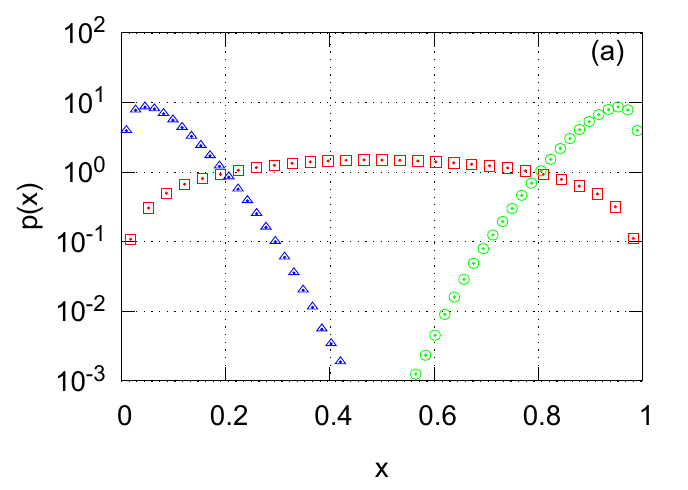}
   \hspace{0.1\textwidth}
    \includegraphics[width=0.4\textwidth]{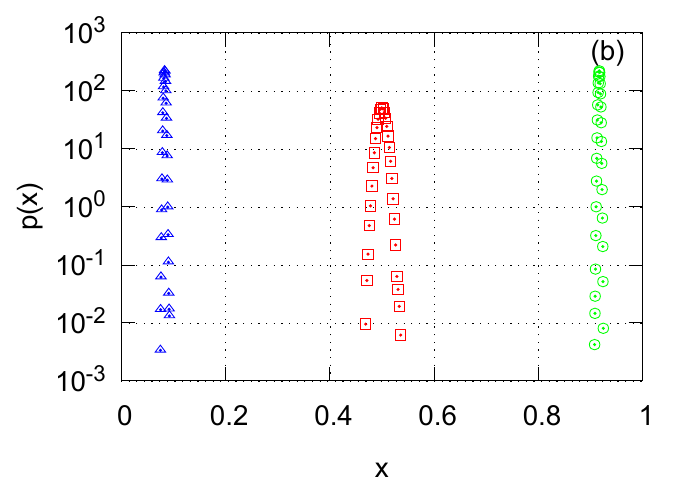}
    \caption{A comparison of a numerically calculated stationary PDF with no controlled agents, $M=M_1=0$ (red squares), and stationary PDF with controlled agents, $M=M_1=20$ (green circles) and $M=20$ and $M_1=0$ (blue triangles), in the non-extensive (a) and extensive (b) case. Model parameters were set as follows: $\sigma_1=\sigma_2=2$, $h=1$. A stochastic model, Eq. (\ref{eq:cnonextsde}), was used for (a) and agent-based model, Eq. (\ref{eq:cext}), with $N=10^4$ was used for (b).}
   \label{fig:pdfs}
\end{figure}

\begin{figure}
    \centering
    \includegraphics[width=0.4\textwidth]{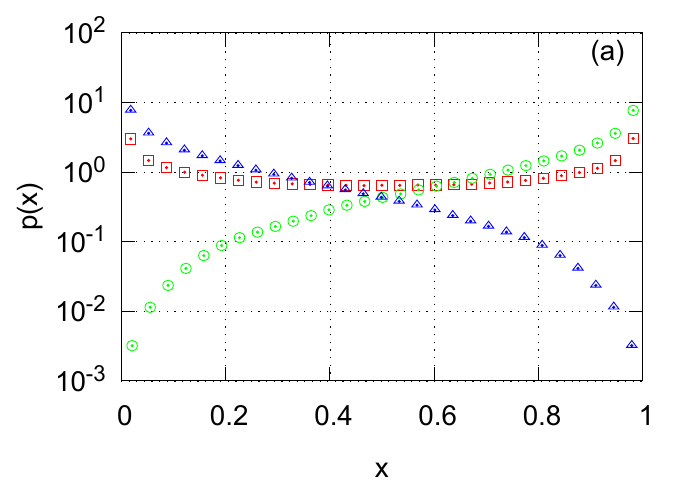}
   \hspace{0.1\textwidth}
    \includegraphics[width=0.4\textwidth]{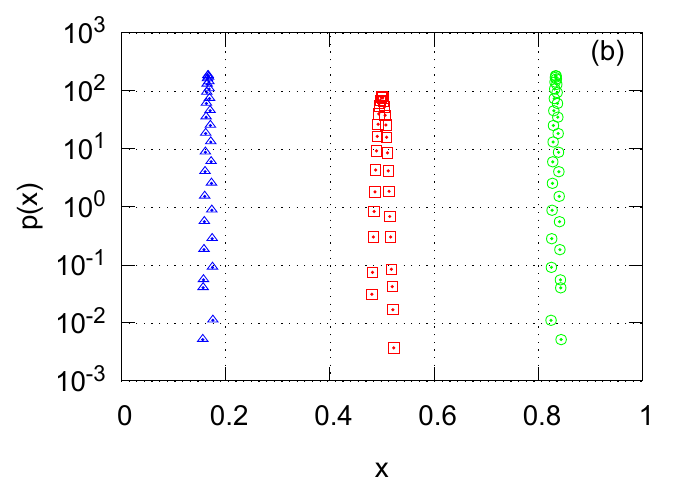}
    \caption{A comparison of a numerically calculated stationary PDF with no controlled agents, $M=M_1=0$ (red squares), and stationary PDF with controlled agents, $M=M_1=2$ (green circles) and $M=2$ and $M_1=0$ (blue triangles), in the non-extensive (a) and extensive (b) case. Model parameters were set as follows: $\sigma_1=\sigma_2=0.5$, $h=1$. A stochastic model, Eq. (\ref{eq:cnonextsde}), was used for (a) and agent-based model, Eq. (\ref{eq:cext}), with $N=10^4$ was used for (b).}
   \label{fig:pdfs-herding}
\end{figure}

In Fig. \ref{fig:pdfs} we numerically confirm that a fixed small number of the controlled agents
($M=20$) enables us to significantly shift the stationary probability density function (abbr. PDF) of the macroscopic variable
 to the desired end despite the fact that the agents have strong individualistic tendencies, $\sigma_i > h$. If the agents have
stronger herding behavior tendencies, $\sigma_i < h$, then the impact of the controlled agents is even stronger. In Fig.
\ref{fig:pdfs-herding} we show that as few as two controlled agents are enough to significantly influence the stationary PDF
of the model if herding behavior is prevalent.

An important question in this context is how fast the controlled agents are able to
make the desired impact. Or namely, how fast the statistical properties of the system, PDF and mean,
converge to the stationary ones. In case the other agents interact extensively the answer
can be obtained analytically by solving corresponding ordinary differential equation,  Eq.
(\ref{eq:cextsde}). Its solution is given by:
\begin{equation}
x(t) = x_f + [x(0) - x_f] \exp( -[h M + \sigma_1 + \sigma_2 ] t) ,\label{eq:extsol}
\end{equation}
here $x(0)$ is the initial condition and $x_f$ is a fixed point of Eq. (\ref{eq:cextsde}), which is given
by Eq. (\ref{eq:exmean}).

\begin{figure}
    \centering
    \includegraphics[width=0.4\textwidth]{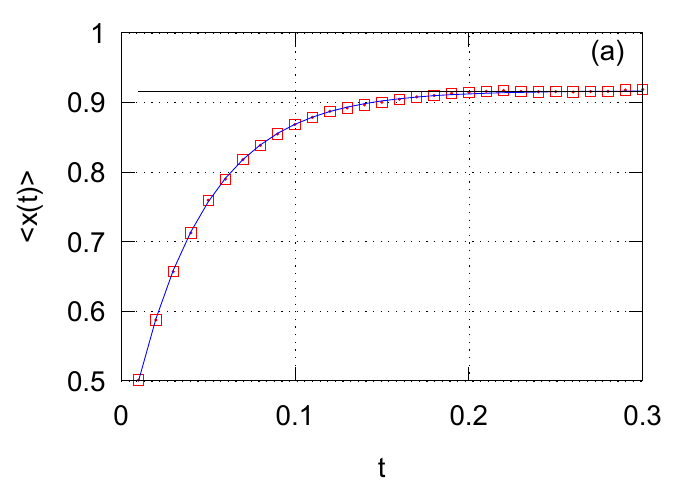}
   \hspace{0.1\textwidth}
    \includegraphics[width=0.4\textwidth]{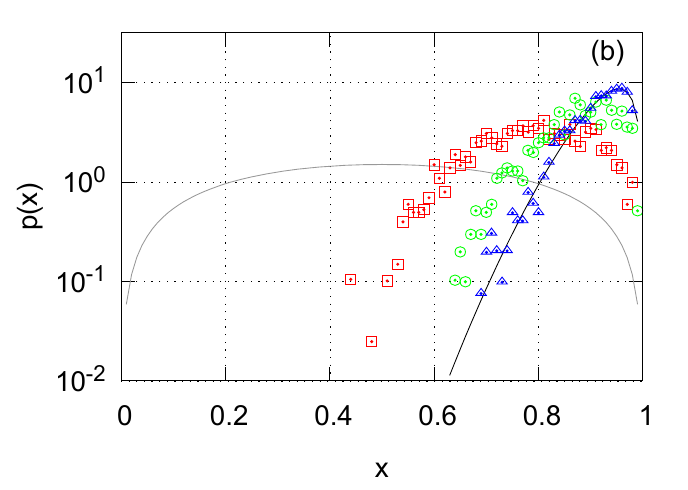}
    \caption{Time evolution of the mean (a) and the PDF (b) of $1000$ time series obtained by
numerically solving the non-extensive stochastic model, Eq. (\ref{eq:cnonextsde}). Subfigure
(a): red squares represent the mean trajectory (average over ensemble of $1000$ realizations),
blue curve is a plot of Eq. (\ref{eq:extsol}), while black curve represents the expected mean.
Subfigure (b): different types of points represent PDF snapshots at distinct times (red squares -
$t=0.05$, green circles - $t=0.1$, blue trinagles - $t=0.15$), gray curve represents the initial
condition (the PDF at $t=0$), while black curve represents the expected PDF. Model parameters
were set as follows: $\sigma_1 = \sigma_2 =2$, $h=1$, $M=M_1=20$.}
   \label{fig:convergence}
\end{figure}

It is a more complex task to solve the non-extensive case, Eq. (\ref{eq:cnonextsde}). One has to
find the eigenvalues of the Fokker-Planck equation \cite{Risken1996Springer}. The problem is that
the corresponding Fokker-Planck equation appears to be too
complex to be dealt with analytically. A viable alternative, of course, is a numerical simulation.
In Fig. \ref{fig:convergence} we plot the results of a numerical simulation, which show that
the convergence times are finite for both mean and stationary PDF. Furthermore the obtained results show
that the convergence of the mean is well described by the Eq. (\ref{eq:extsol}), and thus the
convergence in both cases happen exponentially fast.

\section{Conclusions}
\label{sec:conclusion}

In this contribution we have approached modeling of the control of collective behavior in complex
socio-economic systems. Namely, we have modified a well-known agent-based herding model (originally
introduced in \cite{Kirman1993QJE}) to include agents, whose state is controlled exogenously. The control over
a small number of the agents enabled us to significantly influence the behavior of the other agents,
who act based on the rules of the original agent-based herding model. We found that even an infinitely
large systems would be strongly affected, if the controlled agents interact with the other agents on the global scale
(non-extensively). We find this to be in agreement with the related experiments \cite{Dyer2009RSTB,Krause2011TEcoEvo}. 

The control is even more effective if the other agents interact among themselves on the local scale
(extensively). This may correspond to the actual dynamics of real societies, where most of the people
interact in their ``neighborhood`` (i.e. friends, coworkers and other acquaintances), while there are also
some prominent individuals, leaders, who are able to influence the society on a global scale. Thus the
mathematical model presented in this contribution may be also considered as a microscopic explanation
of the leadership phenomenon in the contemporary society. The leadership, in this sense, can be seen
as an exogenous information field in the similar way as parameters $\sigma_i$ can.

The ability to control the macroscopic dynamics of the agent-based herding model may allow future
developments of the tools for preventing financial bubbles or crashes. It is thought that these extreme
events may be caused by the herding behavior \cite{Preis2012SciRep,Parisi2013PhysRevE}, which is also
the main ingredient of the model allowing control of the macroscopic dynamics. Furthermore the considered
agent-based herding model was previously used to construct a simple agent-based models for the financial
markets \cite{Kononovicius2012PhysA,Alfarano2005CompEco,Alfarano2008Dyncon,Kononovicius2013EPL}.
In these models some of the states available to agents increase the market volatility (these states are related
to the chartist trading strategies), while the others decrease (these states are related to the fundamentalist
trading strategies). By using the results obtained in this contribution one could attract more agents to the lower volatility
state thus decreasing volatility of the whole market.


\end{document}